# Local stress in cylindrically curved lipid membrane: insights into local versus global lateral fluidity models

**Konstantin V. Pinigin**

A. N. Frumkin Institute of Physical Chemistry and Electrochemistry, Russian Academy of Sciences, 31/4 Leninskiy Prospekt, 119071 Moscow, Russia

Correspondence: piniginkv@gmail.com;

**Abstract:** Lipid membranes, fundamental to cellular function, undergo various mechanical deformations. Accurate modeling of these processes necessitates a thorough understanding of membrane elasticity. The lateral shear modulus, a critical parameter describing membrane resistance to lateral stresses, remains elusive due to the membrane's fluid nature. Two contrasting hypotheses, local fluidity and global fluidity, have been proposed. While the former suggests a zero local lateral shear modulus anywhere within lipid monolayers, the latter posits that only the integral of this modulus over the monolayer thickness vanishes. These differing models lead to distinct estimations of other elastic moduli and affect the modeling of biological processes, such as membrane fusion/fission and membrane-mediated interactions. Notably, they predict distinct local stress distributions in cylindrically curved membranes. The local fluidity model proposes isotropic local lateral stress, whereas the global fluidity model predicts anisotropy due to anisotropic local lateral stretching of lipid monolayers. Using molecular dynamics simulations, this study directly investigates these models by analyzing local stress in a cylindrically curved membrane. The results conclusively demonstrate the existence of a static local lateral shear stress and anisotropy in local lateral stress within the monolayers of the cylindrical membrane, strongly supporting the global fluidity model. These findings have significant implications for the calculation of surface elastic moduli and offer novel insights into the fundamental principles governing lipid membrane elasticity.

# 1. Introduction

Lipid membranes, essential components of living organisms, serve as semipermeable barriers for cells and their organelles [1]. Beyond their structural role, they are integral to vital cellular processes such as vesicle-mediated transport [2,3], cell division [4], viral budding [5,6], membrane protein sorting [7–9], membrane-mediated interactions [10–15], and lipid-protein interactions [16]. These processes involve mechanical deformations of the membrane, necessitating a theoretical understanding of membrane elasticity to accurately describe their energetics and dynamics.

Classical elasticity theory models lipid membranes as continuous, laterally fluid, elastic 3D bodies [14,17–21]. The free energy of deformation is governed by elastic parameters characterizing the energy cost of specific deformation modes. Assuming membrane incompressibility, three independent local elastic moduli—stretching, transverse shear, and lateral shear modulus—describe these deformations [14,17,20]. While the local stretching modulus is relatively well-known [22], the others remain less characterized. The transverse shear modulus influences lipid tilt and tilt modulus, relevant for deformations on the scale of membrane thickness [14,17]. The value of the transverse shear modulus can be reasonably theoretically estimated from the consideration of the oil-water interface [17].

The lateral shear modulus, representing the membrane's resistance to lateral deformation, is particularly challenging to incorporate into elasticity theory due to the lateral mobility of lipid molecules. Two primary assumptions—local and global fluidity—have been proposed to address this challenge [17]. The local fluidity assumption posits a zero lateral shear modulus throughout the lipid monolayers, while the global fluidity assumption requires only the integral of the lateral shear modulus over the monolayer thickness to be zero. Some studies employ the local fluidity assumption [18,19,23–25], while others consider the possibility of global fluidity [14,17,20,26–29]. The choice between these assumptions significantly impacts the derived surface elastic moduli, including bending, Gaussian curvature, and twist moduli [14,17,20].

Previous attempts to indirectly determine the correct fluidity assumption have yielded conflicting results. Theoretical analyses have suggested that the global fluidity assumption implies local fluidity due to the classical stability requirements [14], while molecular dynamics observations of membrane undulations have indicated a nonzero twist modulus [20,28,29], implying a non-zero local lateral shear modulus. However, these molecular dynamics findings may be influenced by limitations of the elasticity theory at short undulation wavelengths [14].

This work aims to directly determine the correct fluidity assumption using molecular dynamics simulations. In Section 3, it is shown that by analyzing the local stress tensor in cylindrically curved lipid bilayers, it is possible to distinguish between local and global fluidity models. The md simulation results, presented in Section 4, provide insights into the nature of membrane fluidity and its implications for understanding membrane deformation processes.

# 2. Materials and Methods

To conduct the molecular dynamics simulations, a coarse-grained force field was selected. Given the conceptual nature of the problem and the focus on studying fluidity properties, a coarse-grained representation of lipid bilayers is sufficient. These simplified models retain the essential lateral fluidity characteristic of lipid membranes, allowing lipids to diffuse within the membrane plane. Moreover, coarse-grained models offer significant computational advantages.

The Martini force field [30], a widely used option for biomolecular simulations, was employed in this study. To further reduce computational costs, the Dry Martini variant [31], which utilizes an implicit solvent model, was chosen. In addition, the use of a wet Martini force may complicate the analysis due to potential coupling of membrane undulations between periodic images of cylindrical bilayers, hindering the accurate interpretation of results.

Palmitoyloleoylphosphatidylcholine (POPC), which is present in living cells [32] and commonly used in model membranes [33] and molecular dynamics simulations [34], was selected to construct the lipid bilayers. In the Dry Martini force field, POPC is represented by 12 beads: a choline bead, a phosphate bead, two glycerol beads, and four beads in each tail.

*2.1 Molecular dynamics parameters*

MD simulations were performed using GROMACS [35,36]. Dry Martini v2.1 [31] simulations were conducted using the stochastic dynamics integrator [37] with a 30 fs time step and a friction constant of 4.0 ps. Non-bonded interactions were calculated using the Verlet algorithm [38] with a 1.4 nm neighbor list updated every 10 steps. The van der Waals and Coulomb cutoff distances were set to 1.2 nm, with van der Waals interactions switched to zero using a force-switch modifier at 0.9 nm. Coulomb interactions employed a reaction-field scheme with a dielectric constant of 15. Prior to the main simulations, the systems were energy-minimized using the steepest descent integrator and soft-core potentials [39], followed by a 0.5 ns equilibration run with a 10 fs time step. The temperature was maintained at 300 K. Dry Martini POPC remains in a fluid state over a wide range of temperatures, exhibiting no phase transition to the gel phase [40].

*2.2 Planar lipid bilayer*

A planar lipid bilayer comprising 256 lipids (128 lipids per monolayer) was generated using the CHARMM-GUI software [41–45]. A semi-isotropic pressure coupling scheme with the Berendsen barostat [46] was employed to maintain a lateral pressure of 0 bar. The lateral box compressibility was set to $3 \times 10^{-4}$ bar$^{-1}$, while the bilayer thickness (z-axis) was fixed at 10.5 nm. Before the production run of 6 $\mu$s, the bilayer was equilibrated for 130 ns.

*2.3 Cylindrical lipid bilayer*

The MD simulation of a cylindrically curved lipid bilayer requires more lipid molecules, which may lead to a large amplitude of thermal undulations of the membranes shape. A thorough analysis of lipid tube undulations was performed in Ref. [47]. In general, smaller radii lead to smaller undulation magnitudes due to the reduced number of molecules. However, at a given tube radius, there exists a tube length $L_0$ where undulations are minimal, while they diverge at $L > L_0$ [47]. The prediction of amplitude divergence, however, is not confirmed by MD simulations [48], showing that undulation amplitude decreases even at $L > L_0$.

The cylindrically curved lipid bilayer was constructed using the BUMPy software [49]. The radius at the bilayer midsurface was set to 6.5 nm To minimize membrane undulations, the tube length was set to 20 nm, which is approximately the optimal length for this radius given a bending rigidity of 20 $k_BT$ (typical for lipid bilayers) and an undulation cutoff of 2 nm (approximately monolayer thickness) [47]. The pivotal plane was positioned at 1 nm, resulting in a lipid tube with 1483 and 1087 lipids in the outer and inner monolayer, respectively.

MD simulations of the lipid tube were conducted in the NVT ensemble, with an initial equilibration run of 30 nanoseconds. The simulation box size was 20 nm along the tube axis and 40 nm in the other two directions. Lipid flip-flop, the movement of lipids between the monolayers, does not occur in Dry Martini lipid bilayers [31]. To balance the lipid distribution between the monolayers, inverted cylindrical flat-bottom position restraints with a force constant of 1000 kJ/mol/nm$^2$ were applied to the hydrocarbon lipid tails to open membrane pores. These restraints acted along axes perpendicular to each other and the tube axis, creating four pores. Initially, the restraint distance was set to 2 nm for 30 ns to facilitate rapid lipid transfer. Subsequently, it was reduced to 1.2 nm and simulated for 300 ns at this value. The distance constants of the position restraints were changed gradually in steps of 0.2-0.5 nm to avoid strong perturbations in the system; each step was simulated for 1.5 ns. This pore-opening approach is similar to those described in Ref. [49–51]. During the time the pores were open, 172 lipids moved from the outer monolayer to the inner monolayer, while 175 lipids moved from the inner monolayer to the outer monolayer, resulting in 1486 and 1084 lipids in the outer and inner monolayers, respectively. Thus, the ratio of lipids in the monolayers did not change significantly, implying that the initial choice of the pivotal plane at 1 nm approximately corresponds to the actual value of the pivotal plane's position. Following restraint removal, the lipid tube was further equilibrated for 100 ns before a 3 $\mu$s production run.

*2.4 Determination of local stress*

During MD simulations, particle coordinates and velocities were recorded every 4.98 ps. The resulting frames were analyzed using GROMACS-LS software [52–55], employing the covariant central force-decomposition for multi-body potentials [53]. Statistical analysis involved block averaging (10 blocks) followed by repeated resampling (200 iterations).

In GROMACS-LS, the stress tensor is calculated using the Irving-Kirkwood-Noll definition [56,57] of local stress and spatial averaging with trilinear weighting functions over a 3D rectangular grid [52]. For the planar bilayer, the grid size was set to 0.05 nm perpendicular to the bilayer and the box size in other two directions. For the cylindrical bilayer, the grid size was the box size along the cylinder axis and 0.05 nm in other two directions. Before analyzing the local stress, at each simulation frame lipid bilayers were spatially translated with the help of MDAnalysis 2.7.0 [58,59] to ensure that their geometric center coincided with the center of the simulation box.

To obtain local stress components in cylindrical coordinates, the calculated local stress tensor, $\Sigma$, was transformed at each grid point using the rotation matrix $A_\theta$ around the cylinder axis by the azimuthal angle $\theta$: $\Sigma' = A_\theta^T \Sigma A_\theta$. The stress components with radial coordinates $r \pm 0.025$ nm (where $r$ ranges from 0.0 nm to 19.95 nm with a step of 0.05 nm; $r = 0.0$ nm corresponds to the cylinder axis passing through the box center) were averaged to determine the radial dependence of the local stress components in cylindrical coordinates.

## 3. Theory

### 3.1 Elastic energy

Within the framework of classical elasticity theory, a lipid monolayer is considered as a continuous, three-dimensional elastic medium [14,17,19,20,22,60]. The average orientation of lipid molecules is characterized by a unit vector field, referred to as the director, that extends from the lipid heads to the lipid tails. The monolayer's planar state serves as a reference configuration. In this state, the director field is perpendicular to the monolayer plane, and the monolayer exhibits transverse isotropy with respect to an axis aligned with the director field. The elastic energy density of the lipid monolayer relative to the reference state should be expressed in a manner consistent with this transverse isotropy. To facilitate this, a Cartesian coordinate system $xyz$ is introduced with the $xy$ plane parallel to the monolayer in its initial configuration and the $z$-axis aligned with the director field. An arbitrary deformation of the monolayer can be described by a vector function $\mathbf{R}(x,y,z)$, which maps the monolayer points with coordinates $(x,y,z)$ to some other points $\mathbf{R}(x,y,z)$. In the chosen coordinate system, the general quadratic expression for the elastic energy density of the transversely isotropic material has the following form [61]:

$$w = \sigma_l \left(u_{xx} + u_{yy}\right) + \sigma_z u_{zz} + \frac{1}{2}\lambda_1 \left(u_{xx} + u_{yy}\right)^2 + \frac{1}{2}\lambda_2 u_{zz}^2 + \lambda_{12} \left(u_{xx} + u_{yy}\right) u_{zz}$$
$$+ \frac{1}{2}\lambda_S \left[\left(u_{xx} - u_{yy}\right)^2 + 4u_{xy}^2\right] + 2\lambda_T \left(u_{xz}^2 + u_{yz}^2\right) \quad (1)$$

, where $u_{ij}$ ($i, j \in \{x, y, z\}$) are the components of the Green-Lagrange strain tensor $\mathbf{U} \equiv \frac{1}{2}\left(\mathbf{F}^T\mathbf{F} - \mathbf{I}\right)$ ($\mathbf{F}$ is the Jacobian matrix of the deformation $\mathbf{R}$); $\lambda_1$, $\lambda_2$, $\lambda_3$, $\lambda_S$, $\lambda_T$ are the elastic moduli; $\sigma_l$ and $\sigma_z$ are the pre-stress terms. Due to transverse isotropy, the elastic moduli and pre-stress terms are independent of $x$ and $y$ but may vary with $z$. Eq. (1) shows that the possible deformation modes can be categorized into six terms: $u_{xx} + u_{yy}$, $u_{zz}$, $u_{xx} - u_{yy}$, $u_{xy}$, $u_{xz}$ and $u_{yz}$. The last two modes represent lipid tilt [14,17,20], or the deviation of the director field from the monolayer normal. In the following, only deformations in which the director field remains perpendicular to the monolayer will be considered, which implies $u_{xz} = u_{yz} = 0$. The first two modes, $u_{xx} + u_{yy}$ and $u_{zz}$, correspond to lateral and longitudinal stretching, respectively, which are characterized by the moduli $\lambda_1$, $\lambda_2$, and the coupling modulus $\lambda_{12}$. The remaining deformation modes, $u_{xx} - u_{yy}$ and $u_{xy}$, represent lateral shear, and are characterized by the lateral shear modulus, $\lambda_S$. If $u_{xx}$, $u_{yy}$ and $u_{xy}$ are coordinate-independent, the $u_{xy}$ term can be eliminated from the elastic energy expression by rotating the Cartesian coordinate system around the $z$-axis. The new coordinate axes should align with the principal axes of the strain ellipsoid in the lateral plane [62].

Given the evidence from experiments [63–65] and MD simulations [66–68], it is common to assume that lipid monolayers are incompressible in terms of volume. Up to the first order in $u_{ij}$, the incompressibility condition can be written as $u_{zz} = -\left(u_{xx} + u_{yy}\right)$. Substituting this condition into Eq. (1) and also assuming that $u_{xz} = u_{yz} = 0$, Eq. (1) simplifies to:

$$w = \sigma_0 \left(u_{xx} + u_{yy}\right) + \frac{1}{2} E \left(u_{xx} + u_{yy}\right)^2 + \frac{1}{2} \lambda_S \left[\left(u_{xx} - u_{yy}\right)^2 + 4 u_{xy}^2\right], \qquad (2)$$

where $\sigma_0 = \sigma_l - \sigma_z$ and $E = \lambda_1 + \lambda_2 - 2\lambda_{12}$.

*3.2 Fluidity assumptions*

Lipid monolayers are characteristically fluid in the lateral direction, enabling lipid molecules to diffuse freely within the plane of the monolayer. This lateral fluidity necessitates specific assumptions regarding the lateral shear modulus, $\lambda_S$, a measure of the resistance of a material to deformation by shear stress. The two most common fluidity assumptions, (i) local fluidity assumption and (ii) global fluidity assumption, were outlined in Ref. [17]. In the following subsections, these assumptions are discussed in detail along with their implications for the local stress tensor.

3.2.1 Local fluidity assumption

Under the local fluidity assumption, it is assumed that the lateral shear modulus, $\lambda_S(z)$, is zero everywhere along the z-axis in the reference configuration of a lipid monolayer [17]:

$$\lambda_S(z) = 0 \quad \text{(local fluidity assumption)}. \qquad (3)$$

This implies that any infinitesimally thin layer of thickness $dz$ within the lipid monolayer, spanning from $z$ to $z + dz$, experiences no resistance to lateral shear deformation. As follows from Eq. (2), the only remaining deformation mode that requires energy is $u_{xx} + u_{yy}$, which is essentially equivalent to the relative lateral area change [17]: $u_{xx} + u_{yy} \approx \varepsilon(z)$, where $\varepsilon(z) = \dfrac{dA'(z)}{dA}$. Here, $dA'(z)$ and $dA$ represent infinitesimal lateral area elements after and before deformation, respectively. Thus, within the local fluidity assumption, the elastic energy is reformulated in terms of $\varepsilon(z)$ [14,17,19]:

$$w_{lf} = \sigma_0(z)\varepsilon(z) + \frac{1}{2} E(z) \varepsilon(z)^2. \qquad (4)$$

Because the elastic energy $w_{lf}$ depends solely on the local stretching $\varepsilon(z)$, the local stress also depends only on $\varepsilon(z)$ [21]. Therefore, since the value of $\varepsilon(z)$ is independent of the coordinate systems, the local stress is laterally isotropic and lacks a lateral shear component [21].

3.2.2 Global fluidity assumption

In contrast to the local fluidity assumption, which stipulates a pointwise zero lateral shear modulus, the global fluidity assumption imposes a less stringent constraint. It requires only that the integral of the lateral shear modulus over the monolayer thickness be zero [17]:

$$\int_{m_0} \lambda_S(z) dz = 0 \quad \text{(global fluidity assumption)}, \qquad (5)$$

where $\int_{m_0} dz$ is the integral over the monolayer thickness in the reference configuration. This assumption is physically grounded in the understanding that lipids can diffuse freely along the monolayer plane only as a collective unit, rather than as individual parts within their thickness. Unlike the local fluidity model, in the global fluidity model, surface elastic parameters, derived from integrating local elastic moduli over the monolayer thickness, depend not only on the local stretching modulus $E(z)$ but also on the local lateral shear modulus $\lambda_S(z)$. Specifically, the combination $\int_{m_0} \lambda_S(z) z^2 \left[\tilde{K}^2 - 4\tilde{K}_G + (\nabla \times \mathbf{T})^2\right]$ should be taken into account [14,17,20], where $\tilde{K}$ is the effective curvature, $\tilde{K}_G$ is the effective Gaussian curvature, $\mathbf{T}$ is the tilt vector and $\nabla \times \mathbf{T}$ is the twist deformation mode. This leads to the correction in the bending modulus (coefficient of $\tilde{K}^2$) and nonzero Gaussian curvature and twist moduli, which are zero in the local fluidity model [14,23].

Under the assumption of global fluidity, the local stress is generally laterally anisotropic. The general expression for the Cauchy stress tensor is given by [69]:

$$\Sigma = \frac{1}{\det \mathbf{F}} \mathbf{F} \frac{\partial w}{\partial \mathbf{U}} \mathbf{F}^T, \tag{6}$$

where $\det \mathbf{F}$ is the determinant of $\mathbf{F}$. It is convenient to express $\mathbf{F}$ and $\mathbf{U}$ in terms of the local basis $\mathbf{e}_i$ of the deformation $\mathbf{R}$: $\mathbf{e}_i = \frac{\partial \mathbf{R}}{\partial x^i}$, where $(x^1, x^2, x^3) = (x, y, z)$. The columns of the matrix $\mathbf{F}$ are the vectors $\mathbf{e}_i$, while $u_{ij} = \frac{1}{2}(\mathbf{e}_i \cdot \mathbf{e}_j - \delta_{ij})$, where $\delta_{ij}$ is the Kronecker delta. Let's consider deformations where $u_{xx}$ and $u_{yy}$ are coordinate-independent, assuming the monolayer remains planar. Then, $\mathbf{e}_1 = \sqrt{1+2u_{xx}}\,\mathbf{i}$, $\mathbf{e}_2 = \sqrt{1+2u_{yy}}\,\mathbf{j}$, where $\mathbf{i}$ and $\mathbf{j}$ are the unit vectors along the $x$ and $y$-axis, respectively. From the incompressibility condition, it follows that $\det \mathbf{F} = 1$ and $\mathbf{e}_3 = \frac{1}{\sqrt{(1+2u_{xx})(1+2u_{yy})}}\,\mathbf{k}$, where $\mathbf{k}$ is the unit vector along the $z$-axis. Using these expressions for $\mathbf{e}_i$ and substituting Eq. (2) into Eq. (6), we obtain:

$$\Sigma_{xx} = \left[\sigma_0(z) + E(z)(u_{xx} + u_{yy}) + \lambda_S(z)(u_{xx} - u_{yy})\right](1 + u_{xx}) + P,$$

$$\Sigma_{yy} = \left[\sigma_0(z) + E(x)(u_{xx} + u_{yy}) - \lambda_S(z)(u_{xx} - u_{yy})\right](1 + u_{yy}) + P, \tag{7}$$

$$\Sigma_{zz} = P,$$

$$\Sigma_{xy} = \Sigma_{yx} = \Sigma_{xz} = \Sigma_{zx} = \Sigma_{yz} = \Sigma_{zy} = 0,$$

where $\Sigma_{ij}$ are the components of the stress tensor and $P$ is the Lagrange multiplier introduced to account for the incompressibility constraint. From Eqs. (7) it follows that, the difference $\Sigma_{xx} - \Sigma_{yy}$ can be written as:

$$\Sigma_{xx} - \Sigma_{yy} = \left[\sigma_0(z) + E(z)(u_{xx} + u_{yy}) + \lambda_S(z)(2 + u_{xx} + u_{yy})\right](u_{xx} - u_{yy}). \tag{8}$$

Since the right-hand side of Eq. (8) is generally non-zero, this indicates that, under the assumption of global lateral fluidity, the lateral stress is generally anisotropic.

It is important to note that the idealized state of a flat monolayer, where it is stretched in one direction and compressed in the other, is somewhat hypothetical. In reality, lipid molecules would rapidly reorganize to achieve a more stable laterally isotropic state and uniform lateral stress. However, specific geometric constraints can hinder this relaxation process.

Consider, for instance, the deformation of a planar monolayer into a cylindrical shape. Without loss of generality, we can assume that the monolayer is bent around the $y$-axis. In this scenario, the $yy$ component of the strain tensor ($u_{yy}$) is independent of coordinates, while the $xx$-components ($u_{xx}$) varies with $z$. Consequently, in the cylindrical configuration, even if lipids reorganize through lateral movement (assuming global lateral fluidity), the difference between $u_{xx}$ and $u_{yy}$ cannot be entirely eliminated along the thickness of the monolayer.

To analyze the local stress in a cylindrical geometry, it is convenient to introduce a cylindrical coordinate system $r\theta z$. Here, $z$ is the axial coordinate along the cylinder's axis, while $r$ and $\theta$ represent the radial distance and azimuth, respectively. In this coordinate system, the stress tensor components are the same as those in Eq. (7):

$$\Sigma_{\theta\theta} = \left[\sigma_0(z) + E(z)(u_{xx} + u_{yy}) + \lambda_S(z)(u_{xx} - u_{yy})\right](1 + u_{xx}) + P,$$

$$\Sigma_{zz} = \left[\sigma_0(z) + E(x)(u_{xx} + u_{yy}) - \lambda_S(z)(u_{xx} - u_{yy})\right](1 + u_{yy}) + P, \tag{9}$$

$$\Sigma_{rr} = P,$$

$$\Sigma_{\theta z} = \Sigma_{z\theta} = \Sigma_{\theta r} = \Sigma_{r\theta} = \Sigma_{zr} = \Sigma_{rz} = 0,$$

where the Lagrange multiplier $P$ is subject to the condition of mechanical equilibrium that the divergence of the stress tensor must vanish: $\frac{\partial}{\partial r}(r\Sigma_{rr}) = \Sigma_{\theta\theta}$. From Eqs. (9) it follows that the difference between the axial component of the local stress ($\Sigma_{zz}$) and the azimuthal component ($\Sigma_{\theta\theta}$) is equivalent to Eq. (8) and is therefore generally non-zero.

Consequently, the local fluidity model and global fluidity model predict different results for local stress in cylindrical lipid monolayers. According to the local fluidity model, lateral stress is always isotropic, even in cylindrical configurations. However, the global fluidity model suggests that local stress should be anisotropic in cylindrical configurations due to the anisotropic lateral stretching of lipid monolayers characteristic of this geometry. Therefore, measuring local stress in cylindrical monolayers can help determine the validity of these models.

## 4. Results

### 4.1 Planar lipid bilayer

This section presents the results of molecular dynamics (MD) simulations performed on the planar POPC bilayer. Figure 1 illustrates the obtained stress profiles as a function of coordinate along the bilayer thickness. The lateral stress profile (Figure 1a) is similar to that reported in Ref. [31]. Three distinct peaks are observed: in the middle of the bilayer, in the transition zone between the hydrophobic and hydrophilic regions, and in the head-group region. These peaks reflect the repulsive nature of interactions in the hydrophobic and head-group regions and the attractive nature of interactions in the transition zone. The lateral stress profile is symmetric with respect to the bilayer midsurface, which is defined as the average position of the terminal beads of lipid tails in this and subsequent sections.

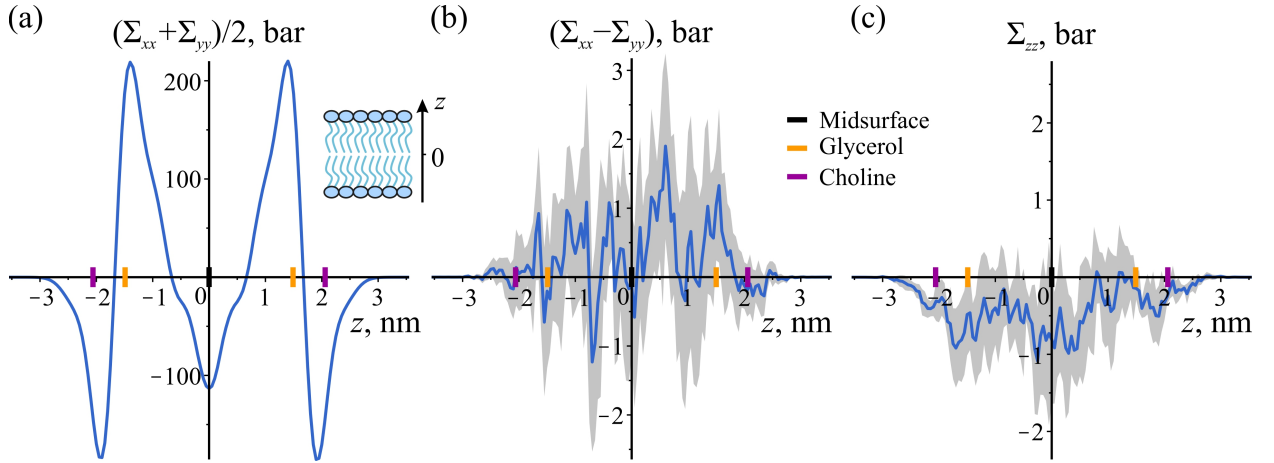

Figure 1. Local stress in planar POPC bilayer. (a) Lateral stress profile. (b) Difference between the lateral components of local stress. (c) Normal component of local stress. The shaded areas represent 95% confidence bands. The shading for the lateral stress profile is omitted due to its small magnitude (less than 2 bar). The inset in panel (a) shows a schematic drawing of a planar bilayer with the $z$-axis indicated. To illustrate the location of lipid monolayers and lipid groups, the enlarged black, orange, and purple tick marks on the $z$-axes are depicted, corresponding to the average positions of the bilayer's midsurface, glycerol, and choline groups of lipids, respectively.

The difference between the lateral stress components, as shown in Figure 1b, is essentially zero within the error bands. This is expected due to the lateral fluidity and transverse isotropy of the bilayer. The normal stress (Figure 1c) is also close to 0 bar, as anticipated. The stability requirement for implicit solvent planar lipid bilayers demands a constant normal stress of zero. A slight systematic shift towards negative values of approximately −0.5 bar may be attributed to a finite time step, leading to minor misconvergence of local stress.

### 4.2 Cylindrically curved lipid bilayer

A cylindrically curved POPC bilayer of radius 6.5 nm and length 20 nm consisting of 2570 POPC lipids was constructed and equilibrated as described in Section 2.3. The local stress components, initially measured in the Cartesian coordinate system, were transformed to the corresponding compo-nents in the cylindrical coordinate system as described in Section 2.4. Figure 2 shows the dependence of the diagonal stress components on the radial coordinate *r*. As anticipated, the stress profiles exhibit asymmetry about the bilayer center, reflecting the differing curvatures of the inner and outer monolayers. Additionally, the azimuthal and axial components of the local stress are not identical, as observed in the figure.

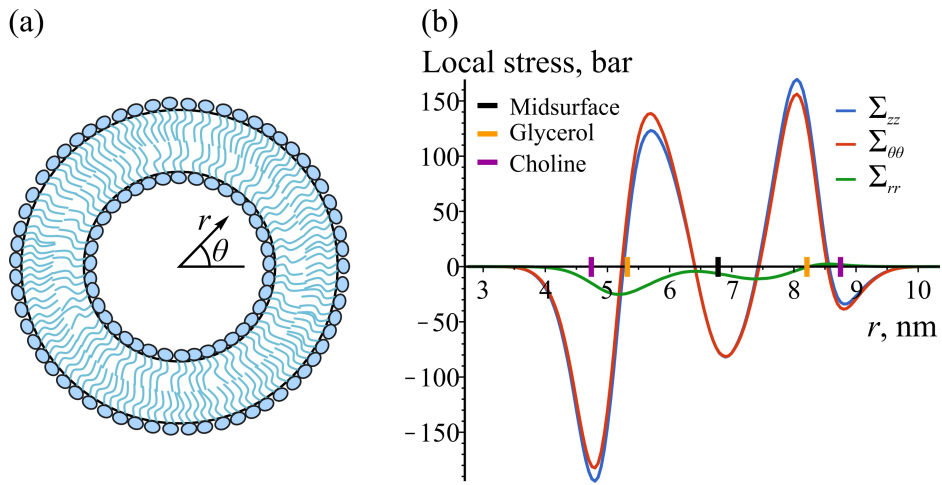

Figure 2. Local stress in cylindrically curved POPC bilayer. (a) A schematic drawing of the cross-section of a cylindrical bilayer, showing the radial coordinate *r* and azimuthal angle *θ*. The axial coordinate *z* is assumed to be perpendicular to the drawing (not shown). (b) The diagonal stress components as functions of the radial coordinate *r*: $\Sigma_{zz}$ (blue), $\Sigma_{\theta\theta}$ (red) and $\Sigma_{rr}$ (green) correspond to the axial, azimuthal and radial component of local stress, respectively. The largest 95% confidence error is approximately 2 bar and is omitted due to its small size. To illustrate the location of lipid monolayers and lipid groups, the enlarged black, orange, and purple tick marks on the *z*-axes are depicted, corresponding to the average positions of the bilayer's midsurface, glycerol, and choline groups of lipids, respectively.

Figure 3 compares the azimuthal ($\Sigma_{\theta\theta}$) and axial ($\Sigma_{zz}$) components of the stress tensor. The difference between these components is substantial within both monolayers, deviating significantly from zero. In the lower monolayer, $\Sigma_{\theta\theta} - \Sigma_{zz}$ is positive, while it is negative in the upper monolayer. At the bilayer's midsurface, the difference becomes zero. For both monolayers, the magnitude of $\Sigma_{\theta\theta} - \Sigma_{zz}$ increases monotonically from the lipid heads, reaching a maximum approximately halfway through the monolayer and then decreasing to zero at the midsurface. The maximum deviation from zero is 17.2 ± 0.7 bar and −13.4 ± 0.3 bar (95% confidence intervals) for the inner and outer monolayer, respectively. These findings indicate that the local stress tensor within both the inner and outer monolayers of these bilayers is laterally anisotropic.

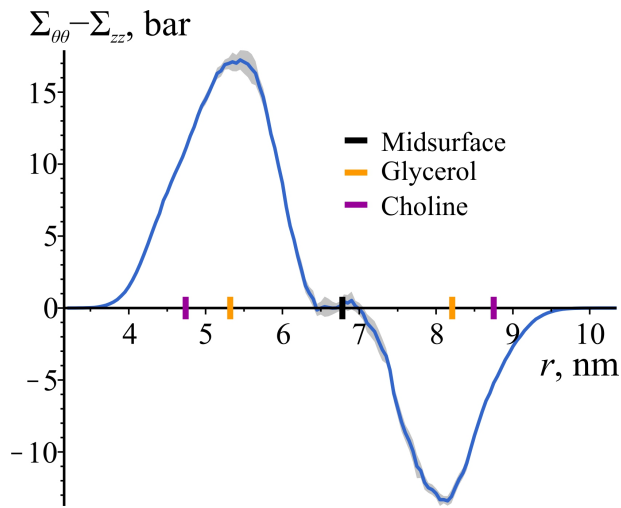

Figure 3. The difference between the azimuthal ($\Sigma_{\theta\theta}$) and the axial ($\Sigma_{zz}$) components of the local stress as a function of the radial coordinate *r* in the cylindrically curved POPC bilayer. The shaded area represents 95% con-

fidence band. To illustrate the location of lipid monolayers and lipid groups, the enlarged black, orange, and purple tick marks on the z-axes are depicted, corresponding to the average positions of the bilayer's midsurface, glycerol, and choline groups of lipids, respectively.

The observed difference in the sign of $\Sigma_{\theta\theta} - \Sigma_{zz}$ between the inner and outer monolayers is likely due to the fact that they are curved in opposite directions, resulting in different strain distributions within the monolayers. $\Sigma_{\theta\theta} - \Sigma_{zz}$ should depend on the local lateral shear modulus, which is challenging to determine by analyzing local stress in cylindrically curved lipid bilayers due to the interaction between the monolayers and its influence on the stress profiles. By removing either the inner or outer monolayer from the saved simulation trajectory, individual contributions to $\Sigma_{\theta\theta} - \Sigma_{zz}$ from each monolayer, as well as from the monolayer-monolayer interaction, can be calculated. As illustrated in Figure 4, the interaction term extends more than 1 nm from the midsurface towards the monolayers. Within each monolayer, $\Sigma_{\theta\theta} - \Sigma_{zz}$ changes sign within the hydrophobic region and also extends approximately 1 nm towards the opposite monolayer.

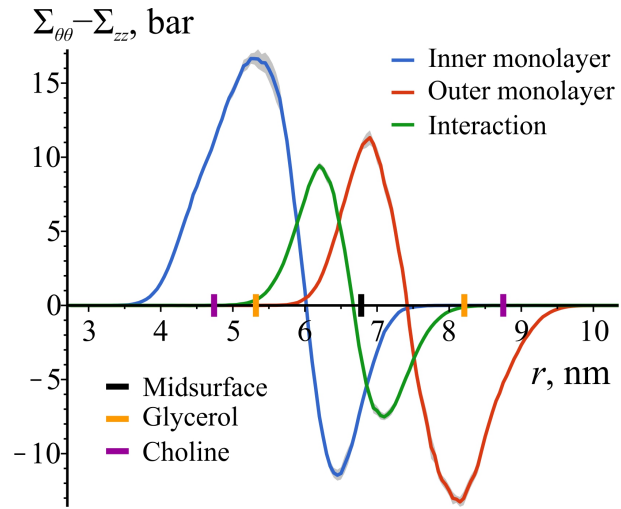

Figure 4. Decomposition of $\Sigma_{\theta\theta} - \Sigma_{zz}$ into the contributions from the inner monolayer (blue), outer monolayer (red) and monolayer-monolayer interaction (green). The shaded areas represent 95% confidence bands. To illustrate the location of lipid monolayers and lipid groups, the enlarged black, orange, and purple tick marks on the z-axes are depicted, corresponding to the average positions of the bilayer's midsurface, glycerol, and choline groups of lipids, respectively.

Due to the cylindrical symmetry of the bilayer under consideration, the off-diagonal components of the local stress tensor should be zero. As illustrated in Figure 5, the off-diagonal component $\Sigma_{z\theta}$ is essentially zero within the error band. Other off-diagonal components exhibit a similar dependence on r and are not shown.

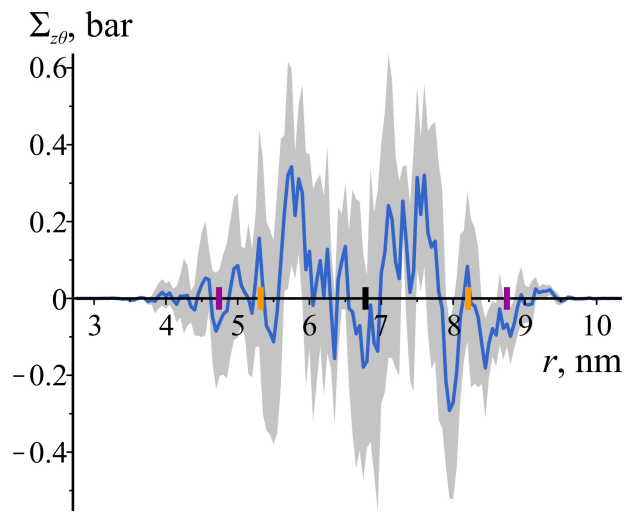

Figure 5. $\Sigma_{z\theta}$ as a function of the radial coordinate $r$ in the cylindrically curved POPC bilayer. The shaded area represents 95% confidence band. To illustrate the location of lipid monolayers and lipid groups, the enlarged black, orange, and purple tick marks on the z-axes are depicted, corresponding to the average positions of the bilayer's midsurface, glycerol, and choline groups of lipids, respectively.

Given that the off-diagonal components of the stress tensor are zero, the axial, azimuthal, and radial directions correspond to the principal directions of the stress tensor. Consequently, the lateral shear stress of maximum magnitude occurs in a coordinate system where one axis aligns with the radial direction, and the other two axes are rotated by $\pm\pi/4$ relative to the axial and azimuthal principal directions. The lateral shear stress corresponding to a rotation by $-\pi/4$ is shown in Figure 6. Its value equals $(\Sigma_{\theta\theta} - \Sigma_{zz})/2$, which corresponds to the off-diagonal component of the local stress after a transformation involving a rotation by $-\pi/4$. Hence, Figure 6 demonstrates the capability of lipid monolayers to resist static local lateral shear stresses.

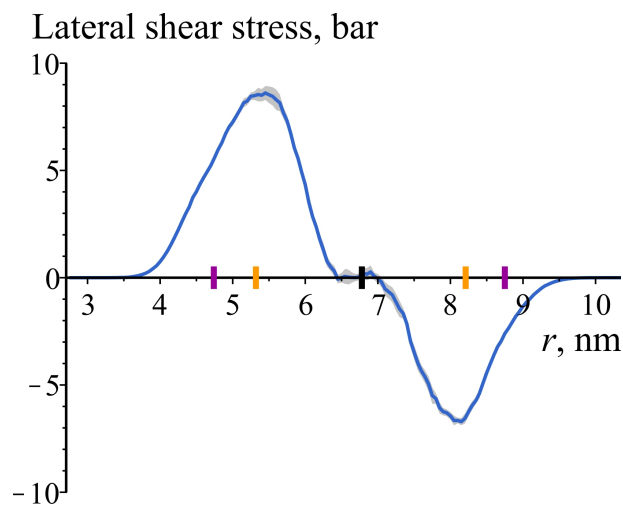

Figure 6. The lateral shear stress of maximum magnitude as a function of the radial coordinate $r$ in the cylindrically curved POPC bilayer. The shaded area represents 95% confidence band. To illustrate the location of lipid monolayers and lipid groups, the enlarged black, orange, and purple tick marks on the z-axes are depicted, corresponding to the average positions of the bilayer's midsurface, glycerol, and choline groups of lipids, respectively.

Section 3 established the mechanical equilibrium condition for the radial and azimuthal stress components: $\frac{\partial}{\partial r}(r\Sigma_{rr}) = \Sigma_{\theta\theta}$. This condition ensures that the bilayer remains structurally stable under the influence of bending stresses. To verify the consistency of the obtained results with this equilibrium condition, Figure 7 illustrates the difference between $\frac{\partial}{\partial r}(r\Sigma_{rr})$ and $\Sigma_{\theta\theta}$, with the derivative $\frac{\partial}{\partial r}(r\Sigma_{rr})$ approximated using the central difference formula. Figure 7 confirms that the equilibrium condition is generally satisfied within the error bounds.

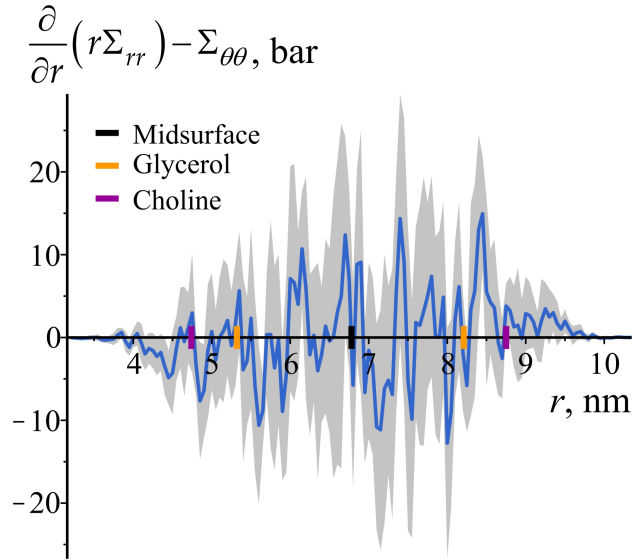

Figure 7. The difference between $\frac{\partial}{\partial r}(r\Sigma_{rr})$ and $\Sigma_{\theta\theta}$ as a function of the radial coordinate $r$ in the cylindrically curved POPC bilayer. The shaded area represents 95% confidence band. To illustrate the location of lipid monolayers and lipid groups, the enlarged black, orange, and purple tick marks on the $z$-axes are depicted, corresponding to the average positions of the bilayer's midsurface, glycerol, and choline groups of lipids, respectively.

## 5. Discussion

The obtained results of molecular dynamics simulations provide compelling evidence supporting the global fluidity assumption for lipid membranes. The analysis of the local stress tensor in the cylindrically curved lipid bilayer revealed that lipid monolayers can resist a static lateral shear stress, indicating a non-zero local lateral shear modulus. This finding directly contradicts the local fluidity assumption, which postulates a zero lateral shear modulus throughout the monolayer thickness. Instead, the results support the global fluidity assumption, where only the integral of the lateral shear modulus over the monolayer thickness is constrained to zero.

The global fluidity assumption has important consequences for the calculation of surface elastic moduli. Unlike the local fluidity assumption, which leads to surface elastic moduli dependent solely on the local stretching and transverse shear moduli, the global fluidity assumption introduces additional contributions from the local lateral shear modulus. This can affect the values of the bending, Gaussian curvature, and twist moduli, which are crucial for describing membrane deformations [14,17,19,20,60]. The value of the local lateral shear modulus is particularly influential on the Gaussian curvature and twist moduli. Within the local fluidity model, the Gaussian curvature and twist moduli are zero, whereas within the global fluidity model, these moduli are generally nonzero [14,23]. The Gaussian curvature modulus is essential for describing fusion/fission events of lipid membranes [70], while the twist modulus is employed in describing membrane-mediated interactions [13,71].

The findings of this work are consistent with previous studies that have indirectly suggested the global fluidity assumption [20,28]. While theoretical analyses have previously indicated that the global fluidity assumption implies local fluidity [14], the direct simulation results provide strong empirical support for the global fluidity model. This suggests that the lateral shear modulus should be negative at some points along the monolayer thickness, contradicting classical stability requirements [14]. However, the local stretching modulus is also known to be negative in certain regions, especially the head-group region [22]. This indicates the presence of a stabilization mechanism, possibly related to the structural properties of lipid molecules and their director deformations.

Living cells contain tubular membrane structures, such as those found in the endoplasmic reticulum [72,73] and tunneling nanotubes [74]. Additionally, neck-like structures form during fusion and fission events [75,76], which, like cylinders, exhibit different principal curvatures and should therefore have anisotropic local lateral stress. The presence of a static lateral shear stress in these structures may influence the functionality and spatial orientation of embedded proteins.

Future studies could aim to directly measure the local lateral shear modulus. However, this is challenging due to the strong coupling between monolayers in cylindrically curved lipid bilayers.

Considering, for example, cylindrically curved lipid monolayers at the oil-water interface might mitigate this difficulty and enable a more direct measurement of the local lateral shear modulus.

**Funding:** The work was supported by the Ministry of Science and Higher Education of the Russian Federation (project No. 122011300058-3).